\newcommand{\be}{\begin{equation}}\newcommand{\ee}{\end{equation}}
\newcommand{\bea}{\begin{eqnarray}}\newcommand{\eea}{\end{eqnarray}}
\newcommand{\nn}{\nonumber\\}\newcommand{\p}[1]{(\ref{#1})}
 \newcommand{\lb}[1]{\label{#1}}
\newcommand\s{\scriptscriptstyle}

\newcommand{\da}{{\dot{\alpha}}}
\newcommand{\db}{{\dot{\beta}}}

\newcommand\A{{\s A}}

\newcommand\ha{{\hat{a}}}
\newcommand\hb{{\hat{b}}}
\newcommand\hc{{\hat{c}}}
\newcommand\he{{\hat{e}}}

\newcommand\hf{{\hat{f}}}

\documentclass[12pt]{article}
\usepackage{latexsym}
\usepackage{color}
\def\theequation{\arabic{section}.\arabic{equation}}
\begin{document}

\begin{titlepage}

\vfill

\begin{center}
\baselineskip=16pt
\vspace{0.2cm}{\Large\bf $SU(4)$ harmonic superspace and supersymmetric gauge theory}

\vskip 0.6cm {\large \bf  B.M. Zupnik } \vspace{1cm}

{\it Bogoliubov Laboratory of Theoretical Physics, JINR, \\
141980 Dubna, Moscow Region, Russia}\\
\vspace{0.3cm}

 {\tt zupnik@theor.jinr.ru}

\end{center}
\vspace{1cm}

\par

We consider the  harmonic-superspace formalism in the $N=4$ supersymmetry using the
$SU(4)/SU(2)\times SU(2)\times U(1)$ harmonics which was earlier applied to the
abelian gauge theory. The  $N=4$ non-abelian constraints in a standard superspace are
reformulated  as  the harmonic-superspace equations for two basic analytic  superfields:
the independent superfield strength $W$ of a dimension 1 and the dimensionless harmonic gauge
4-prepotential $V$ having the $U(1)$ charge 2. These constraint equations I manifestly depend
on the Grassmann coordinates $\theta$, although they  are covariant under the unusual $N=4$
supersymmetry transformations. We analyze an alternative harmonic formalism of the supergauge theory
for two unconstrained nonabelian analytic superfields $W$ and $V$. The gauge-invariant action
$A(W,V)$ in this formalism contains $\theta$ factors in each term, it is invariant under the
$SU(4)$ automorphism group. In this model, the interaction of two infinite-dimensional $N=4$
supermultiplets with the physical and auxiliary fields arises at the level of component fields.
The action $A(W,V)$ generate analytic equations of motion II alternative to the
harmonic-superspace superfield constraints I. Both sets of equations give us the equivalent
equations for the physical component fields of the $N=4$ gauge supermultiplet,
they connect auxiliary and physical fields of two superfields. The nonlinear effective
interaction of the abelian harmonic superfield $W$ is constructed.

\vspace{1cm}


\vfill \vfill \vfill \vfill \vfill
\end{titlepage}

\setcounter{footnote}{0}

\setcounter{page}{1}

\setcounter{equation}0
\section{Introduction}
The superfield constraints for the $ N=3,4, D=4$ gauge theories were considered
in the corresponding superspaces with the coordinates $x^m,~\theta^\alpha_k,
~\bar\theta^{k\dot\alpha}$ \cite{So}, where $k$ describes spinor representations of the $SU(3)$
or $SU(4)$ automorphism groups of these supersymmetries.
It was shown that the constraints for superfield strengths yield  the equations
of motion.

We use the harmonic superspace (HS) approach to the supersymmetric gauge theory and supergravity
described  in the book  \cite{GIOS}. The important achievement of this approach is the off-shell
superfield formalism of the $ N=3$ supersymmetric Yang-Mills theory   using three conjugated
pairs of the $SU(3)$  harmonics $u^1_i,~u^2_i,~u^3_i$ and $u_1^i,~u_2^i,~u_3^i$ describing
the coset $SU(3)/U(1)\times U(1)$\cite{GIK2}. The basic Grassmann-analytic gauge
superfields of this formalism $V^1_2,~V^2_3,~V^1_3$ contain an infinite number
of the component off-shell fields. The Chern-Simons-type action in the $N=3$ harmonic superspace
corresponds to the zero-curvature equations of motion for the superfield strengths.
The HS transform in this theory connects equations of motion in the harmonic superspace with
the superfield constraints in the ordinary $ N=3$ superspace.
The  further development of the $ N=3$  theory was considered in \cite{GIO}-\cite{BISZ}.
The classical equations of motion of the $ N=3$ gauge theory possess the $ N=4$
supersymmetry \cite{Zu2004}, however, it is not preserved off-shell.

The harmonic superspaces for the $ N=4$ supersymmetry were studied in many papers
\cite{IKNO}- \cite{BP}.  The paper \cite{Sok}
describes the twistor-harmonic superspace based on the bosonic spinor harmonics of the
Lorentz group by analogy with the twistor integrability conditions for $N\geq 3$ Yang-Mills theory
\cite{Wi}. It is not clear how to connect this nonstandard superfield description
with the off-shell field interactions in the ordinary Minkowski space.
As a rule, different harmonic superfield constructions are used for the
description of the short abelian  $ N=4$ supermultiplet on the mass shell \cite{HH,HW2,Howe97,AFSZ}.
The harmonic superspace of ref.\cite{BLS} uses the $USp(4)$ harmonics which do not
guarantee the manifest $SU(4)$ symmetry. Constrained short superfields of different harmonic
formalisms are used in \cite{BS} for a description of effective nonlinear interactions.
The action of the
$N=4$ supersymmetric theory with central charges in the harmonic $USp(4)$ formalism
was expressed via the constrained superfield strength \cite{BP}.

The coset space ${\cal U}_8=SU(4)/H,~~H=SU(2)\times SU(2)\times U(1)$ and corresponding harmonic
coordinates $u^{+a}_k,~u^{-\ha}_k$ are described in Appendix A. We consider in detail the $SU(4)$
invariant harmonic derivatives  and irreducible harmonic combinations in ${\cal U}_8$
having indices of the $SU(4)$ and $H$ groups. The Appendix B is devoted to the analysis
of the ${\cal U}_8$ harmonic analytic superspace ${\cal H}(4+8|8)$. We analyze  conjugation
rules for harmonics, spinor coordinates, harmonic and spinor derivatives and study the
superconformal transformations in the harmonic superspace. The conventions and
formulas from these appendices are widely used in the paper.

The  harmonic superspace ${\cal H}(4+8|8)$ was applied in \cite{HH,HW2,Howe97} to the
description of the abelian superfield strength $W^{++}$  in the $ N=4$
gauge theory on the mass shell. We analyze the non-abelian generalization of this construction in
sect. 2. Our formalism includes the dimensionless analytic gauge $V^{++a}_\ha$-superfields
(4-prepotential) and independent gauge covariant superfield $W^{++}$ of a dimension 1.
The $SU(4)$ self-duality condition for the superfield strength is solved automatically in
this harmonic formalism. We formulate the superfield constraints I in the harmonic superspace.
The solutions
of these constraints connect $W^{++}$ and $V^{++a}_\ha$ superfields, the corresponding gauge
covariant equations depend manifestly on the Grassmann coordinates. Nevertheless, these equations
are covariant under the unusual $N=4$ supersymmetry transformations. Expanding the constraint
equations I in terms of the field components yields the known field equations of the $ N=4$
Yang-Mills theory, all auxiliary fields vanish or are expressed via the physical fields.

In sect. 3, we consider the alternative harmonic superfield formalism of the nonabelian $N=4$
gauge theory and construct the action $A$ of independent unconstrained analytic superfields
$W^{++}$ and
$V^{++a}_\ha$ which includes only first harmonic derivatives of the gauge superfields
and is invariant under the nonabelian gauge group and the automorphism group $SU(4)$.
The action $A$ is an integral on the analytic  harmonic superspace, it manifestly depends
on the Grassmann coordinates $\theta$  and breaks the $N=4$ supersymmetry. The equations of motion
II of the $A$-model are derived by  varying of the action in the independent analytic superfields
 $W^{++}$ and $V^{++a}_\ha$. The superfield equations II are formally equivalent to some combinations
 of the superfield constraints I from sect. 2 with nilpotent $\theta$ multipliers.

Component fields of the infinite-dimensional $ N=4$ supermultiplets $W^{++}$ and $V^{++a}_\ha$
arise in the Grassmann and harmonic expansions of superfields. The component version of the action
$A$ contains an infinite number of fields. The equations of motion of the $A$-model connect the physical
fields of the $N=4$ gauge supermultiplet with some set of auxiliary fields, an infinite number
of additional auxiliary fields vanish on the mass shell. The $N=4$  supersymmetry is restored
after the exclusion of auxiliary component fields, the corresponding equations for the physical fields
are covariant under the $N=4$  supersymmetry.

Section 4 is devoted to the construction of the $N=4$ supersymmetric nonlinear interactions of
the abelian superfield $W^{++}$ by  analogy with the $N=3$ abelian effective self-interaction
\cite{IZ}. We show that the auxiliary fields of the $W^{++}$ superfield play an important role in
the construction of nonlinear effective interactions of the physical $N=4$ fields which describe
the possible quantum corrections in this model.

In sect.5, we study  the manifestly supersymmetric interaction $S$ of the gauge prepotential
$V^{++a}_\ha$ in the harmonic superspace and try to use $S$ as the action of some gauge model.
We find inconsitencies in  this $S$-model which contains
interactions of the standard $N=4$ gauge supermultiplet with additional scalar, vector, tensor
and spinor fields without necessary  additional gauge symmetries.

\setcounter{equation}0
\section{On-shell harmonic superfields in $ N=4$ gauge theory}
\subsection{$ N=4$ superfield constraints in the central basis}
The superfield constraints of the $ N=4$ Yang--Mills theory are described by the
following equations \cite{So}:
\bea
  &&\{\nabla^k_\alpha,\nabla^j_\beta\} =
\varepsilon_{\alpha\beta}W^{kj}~,\quad
\{\bar\nabla_{k\dot\alpha},\bar\nabla_{j\dot\beta}\}
=\varepsilon_{\da\db}\bar{W}_{kj}~,
\nn
  &&\{\nabla^k_\alpha,\bar\nabla_{j\dot\beta}\} =-2i\delta^k_j
  \nabla_{\alpha\dot\beta}~,\lb{B3}
\eea
where $\nabla$ are the spinor and vector covariant derivatives in the the central basis (CB)
\bea
&&\nabla_\alpha^k=D_\alpha^k+A_\alpha^k(z),\quad \bar\nabla_{k\dot\alpha}
=\bar{D}_{k\dot\alpha}+\bar{A}_{k\dot\alpha}(z),
\nn
&&\nabla_{\alpha\dot\alpha}=\partial_{\alpha\dot\alpha}+A_{\alpha\dot\alpha}(z)\lb{B3b}
\eea
and the $N=4$ spinor derivatives $D_\alpha^k$ and $\bar{D}_{k\dot\alpha}$ are defined in appendix B.
We use    the  superfield strength $W^{kj}$
\bea
&&W^{kj}=\frac12\varepsilon^{\beta\alpha}(D^k_\alpha A^j_\beta+D^j_\beta A^k_\alpha
+\{A^k_\alpha, A^j_\beta\})
\eea
 and   the conjugated CB superfield strength $\bar{W}_{kj}$.
The subsidiary  $SU(4)$ self-duality condition has the form
\be
\bar{W}_{ij}=\frac12\varepsilon_{ijkl}W^{kl}.\lb{Wsd}
\ee
The CB gauge transformations use the anti-Hermitian superfield parameters $C(z)$
\be
\delta A^k_\alpha=-\nabla^k_\alpha C,\quad \delta \bar{A}_{k\dot\alpha}=-\bar\nabla_{k\dot\alpha} C,
\quad \delta W^{kl}=[C,W^{kl}].
\ee

The Bianchi identities of a dimension $3/2$ yield the CB equations of motion for $W^{ij}$
\cite{So}
\bea
&&2\nabla^k_\gamma W^{ij}-\nabla^i_\gamma W^{jk}-\nabla^j_\gamma W^{ki}=0,\nn
&&\bar\nabla_{k\dot\alpha}W^{ij}+\frac13\delta^j_k\bar\nabla_{l\dot\alpha}W^{li}
-\frac13\delta^i_k\bar\nabla_{l\dot\alpha}W^{lj}=0\lb{DW}
\eea
and the conjugated equations for $\bar{W}_{kj}$.

\subsection{Harmonic interpretation of non-abelian superfield constraints}
We use the ${\cal U}_8$ harmonics $u^{+a}_k,~u^{-\ha}_k,~\bar{u}^{-k}_a,~\bar{u}^{+k}_\ha $ from
Appendix A and consider harmonic projections of the $ N=4$ constraints \p{B3}
\bea
  &&\{{\bf D}^{+a}_\alpha,{\bf D}^{+b}_\beta\} =
\frac12\varepsilon_{\alpha\beta}\varepsilon^{ba}{\bf W}^{++}~,\nn
&&
\{\bar{\bf D}^+_{\ha\dot\alpha},\bar{\bf D}^+_{\hb\dot\beta}\}
=\frac12\varepsilon_{\dot\alpha\dot\beta}\varepsilon_{\hb\ha}\bar{\bf W}^{++}
=-\frac12\varepsilon_{\dot\alpha\dot\beta}\varepsilon_{\hb\ha}({\bf W}^{++})^\dagger
=\frac12\varepsilon_{\dot\alpha\dot\beta}\varepsilon_{\hb\ha}{\bf W}^{++}~, \lb{B3h}\\
&&\{{\bf D}^{+a}_\alpha,\bar{\bf D}^+_{\ha\dot\beta}\} =0~,\nonumber
\eea
where
\bea
&&{\bf D}^{+a}_\alpha=u^{+a}_k\nabla^k_\alpha=D^{+a}_\alpha+A^{+a}_\alpha,\nn
&&
\bar{\bf D}^+_{\ha\dot\alpha}=\bar{u}^{+k}_\ha\bar\nabla_{k\dot\alpha}=\bar{D}^+_{\ha\dot\alpha}
+\bar{A}^+_{\ha\dot\alpha},\\
&&
{\bf W}^{++}=\varepsilon_{ab}u^{+a}_ku^{+b}_lW^{kl},\quad \bar{\bf W}^{++}=\varepsilon^{\ha\hb}
\bar{u}^{+k}_\ha\bar{u}^{+l}_\hb\bar{W}_{kl}.
\eea
Using the harmonic properties \p{usdplus} and \p{uplus2} we connect the self-duality
condition \p{Wsd} with the harmonic condition of anti-Hermicity
\bea
[({\bf W}^{++})^A_B]^\sim=-({\bf W}^{++})^B_A,\quad A, B=1,2,\ldots n.
\eea

We analyze the superfield equations for the superfields $A^{+a}_\alpha,~\bar{A}^+_{\ha\dot\alpha}$
and ${\bf W}^{++}$ using   additional harmonic equations
\bea
&&[\partial^{++a}_\ha,{\bf D}^{+b}_\alpha]=0,\quad [\partial^{++a}_\ha, \bar{\bf D}^+_{\hb\dot\alpha}]=0,
\quad [\partial^{++a}_\ha, \partial^{++b}_\hb]=0,\nn
&&\partial^{++a}_\ha {\bf W}^{++}=0,\quad \partial^{++a}_\ha \bar{\bf W}^{++}=0,\lb{CB1}\\
&&{\bf D}^{+a}_\alpha{\bf W}^{++}=\bar{\bf D}^{+}_{\ha\dot\alpha}{\bf W}^{++}\lb{CB2}
\eea
which are evident in the central basis.

The harmonized $ N=4$ constraints \p{B3h} have more complicated form than the
corresponding Grassmann integrability conditions in the $ N=3$
gauge harmonic-superspace theory \cite{GIOS,GIK2} which do not include superfield strengths
\bea
&&u^1_iu^1_j\{\nabla^i_\alpha,\nabla^j_\beta\} =0~,\quad
u^i_3u^j_3\{\bar\nabla_{i\dot\alpha},\bar\nabla_{j\dot\beta}\}
=0~,\quad
u^1_ku^j_3\{\nabla^k_\alpha,\bar\nabla_{j\dot\beta}\} =0.
\eea

The AB spinor derivatives $D^{+a}_\alpha,~\bar{D}^{+}_{\ha\dot\alpha}$ and the harmonic
derivative $D^{++a}_\ha$ of the analytic basis are defined in Appendix B.
In the non-abelian theory, we define the Hermitian analytic gauge 4-prepotential
\bea
&&(V^{++a}_\hb)_A^B=[(V^{++b}_\ha)_B^A]^\sim,\quad D^{+c}_\alpha V^{++a}_\hb=
\bar{D}^{+}_{\hc\dot\alpha} V^{++a}_\hb=0
\eea
in the adjoint representation of the  gauge
group  $SU(n)$
\bea
&&\delta_\lambda V^{++a}_\hb=-D^{++a}_\hb\lambda +[\lambda,V^{++a}_\hb]\lb{deltaV}
\eea
where $\lambda^B_A$ are the  analytical superfield gauge parameters
\be
\lambda^B_A=-(\lambda_B^A)^\sim.
\ee
The pure analytic harmonic covariant derivative is defined off mass shell
\bea
&&{\cal D}^{++a}_\hb=D^{++a}_\hb+V^{++a}_\hb.
\eea
The commutator of pure analytic   covariant harmonic derivatives
\bea
&&[{\cal D}^{++a}_\ha,{\cal D}^{++b}_\hb]=\frac12F^{(+4)ab}\varepsilon_{\ha\hb}
+\frac12\hat{F}^{(+4)}_{\ha\hb}\varepsilon^{ba},
\eea
 is expressed via two dimensionless analytic gauge covariant superfields
\bea
&&\hat{F}^{(+4)}_{\ha\hb}=\varepsilon_{ab}(D^{++a}_\ha V^{++b}_\hb-D^{++b}_\hb V^{++a}_\ha
+[V^{++a}_\ha, V^{++b}_\hb])=\hat{F}^{(+4)}_{\hb\ha},\nn
&&F^{(+4)ab}=\varepsilon^{\hb\ha}(D^{++a}_\ha V^{++b}_\hb-D^{++b}_\hb V^{++a}_\ha
+[V^{++a}_\ha, V^{++b}_\hb])=-(\hat{F}^{(+4)}_{\ha\hb})^\dagger,\lb{Fstr}\\
&&
(D^{++a}_\ha V^{++b}_\hb)^\dagger=-
D^{++a}_\ha V^{++b}_\hb.\nonumber
\eea

We consider the HS transform from the central basis to the analytic basis
\bea
{\bf D}^{+a}_\alpha=e^{-v}\nabla^{+a}_\alpha e^v,\quad \bar{\bf D}^{+}_{\ha\dot\alpha}
=e^{-v}\bar\nabla^{+}_{\ha\dot\alpha} e^v,\quad{\bf W}^{++}=e^{-v}W^{++}e^v,
\eea
where $v(z,u)\,$ is an anti-Hermitian superfield matrix on the mass shell, and  $W^{++}$ is a
harmonic superfield in AB. We use the $N=4$ noncovariant on-shell representation of the spinor
covariant derivatives in the analytic basis
\bea
&&\nabla^{+a}_\alpha=D^{+a}_\alpha-\frac{1}4\theta^{-a}_\alpha W^{++},\quad
\bar\nabla^+_{\ha\dot\alpha}=\bar{D}^+_{\ha\dot\alpha}+\frac{1}4\bar\theta^-_{\ha\dot\alpha}
W^{++}=(\nabla^{+a}_\alpha)^\dagger\lb{nablspin}
\eea
where $W^{++}$ is an  independent analytic anti-Hermitian covariant superfield
\be
\delta_\lambda W^{++}=[\lambda, W^{++}],\quad (W^{++})^\dagger=-W^{++}.
\ee
By analogy with the  CB constraints \p{B3h}, these spinor AB covariant derivatives satisfy
the  constraints
\bea
&&\{\nabla^{+a}_\alpha,\nabla^{+b}_\beta\} =
\frac12\varepsilon_{\alpha\beta}\varepsilon^{ba}W^{++},\quad
\{\bar\nabla^{+}_{\ha\dot\alpha},\bar\nabla^{+}_{\hb\beta}\} =
\frac12\varepsilon_{\dot\alpha\dot\beta}\varepsilon_{\hb\ha}W^{++},\nn
&&\{\nabla^{+a}_\alpha,,\bar\nabla^{+}_{\hb\beta}\} =0
\eea
where we use the linear relations
\bea
 D^{+b}_\beta W^{++}=\bar{D}^+_{\hb\dot\alpha} W^{++}=0,\quad
 D^{+a}_\alpha\theta^{-b}_\beta=-\varepsilon_{\alpha\beta}\varepsilon^{ba},\quad
\bar{D}^+_{\ha\dot\alpha}\bar\theta^-_{\hb\dot\beta}=\varepsilon_{\dot\alpha\dot\beta}
\varepsilon_{\hb\ha} .
\eea

We define the  on-shell harmonic covariant AB derivative
\bea
&&e^v\partial^{++a}_\ha e^{-v}=\nabla^{++a}_\ha =D^{++a}_\ha+V^{++a}_\ha
-\frac{1}4[(\theta^{-a}\theta^+_\ha)+(\bar\theta^-_\ha\bar\theta^{+a})]
W^{++}\lb{nablaon2}
\eea
which contains the analytic prepotential and the Hermitian non-analytic term manifestly depending
on Grassmann coordinates $\theta$ in accordance with representation \p{nablspin}.
We find that  of the harmonic transform in the  $N=4$ gauge theory and the corresponding
harmonic transform in  the  $N=3$ gauge theory \cite{GIOS} have  essentially different structures.

By the analogy with the trivial CB constraints \p{CB1}, we analyze the following AB constraints:
\bea
&&[\nabla^{++a}_\ha,\nabla^{+b}_\alpha]=0,\quad [\nabla^{++a}_\ha,\bar\nabla^+_{\hb\dot\alpha}]=0,\\
&&[\nabla^{++a}_\ha,\nabla^{++b}_\hb]=0
\eea
using the relations
\be
D^{++a}_\ha \theta^{-b}_\alpha=\varepsilon^{ab}\theta^+_{\ha\alpha},\quad
D^{++a}_\ha\bar\theta^-_{\hb\dot\alpha}=-\varepsilon_{\ha\hb}\bar\theta^{+a}_{\dot\alpha}.
\ee

The nonlinear consistency equations for these AB constraints have the form
\bea
&&I)\quad\mbox{dim.}\,1:\quad {\cal D}^{++a}_\ha W^{++}=D^{++a}_\ha W^{++}+[V^{++a}_\ha, W^{++}]=0,
\lb{hWcon}\\
&&\mbox{dim.}\,0:\quad E^{(+4)ab}=F^{(+4)ab}+(\bar\theta^{+a}\bar\theta^{+b})W^{++}=0,\nn
&&
\hat{E}^{(+4)}_{\ha\hb}=\hat{F}^{(+4)}_{\ha\hb}+(\theta^+_\ha\theta^+_\hb)W^{++}=0,\lb{hcon3}
\eea
and the last equations explicitly depend  on the Grassmann coordinates.
These equations connect two analytic superfield strengths of different dimensions
on the mass shell. The constraint equations are covariant under the nonstandard
$N=4$ supersymmetry transformations
\bea
&&\delta_\epsilon W^{++}=[(\epsilon_k Q^k)+(\bar\epsilon^{k}\bar{Q}_{k})]W^{++},\\
&&\delta_\epsilon V^{++a}_\ha=[(\epsilon_k Q^k)+(\bar\epsilon^{k}\bar{Q}_{k})]V^{++a}_\ha
+\frac12[(\epsilon^{-a}\theta^+_\ha)+(\bar\epsilon^-_\ha\bar\theta^{+a})]W^{++}
,\\
&&\delta_\epsilon \hat{F}^{(+4)}_{\ha\hb}=\varepsilon_{ab}({\cal D}^{++a}_\ha\delta_\epsilon V^{++b}_\hb
-{\cal D}^{++b}_\hb\delta_\epsilon V^{++a}_\ha),\nn
&&\delta_\epsilon \hat{E}^{(+4)}_{\ha\hb}=[(\epsilon_k Q^k)+(\bar\epsilon^{k}\bar{Q}_{k})]
\hat{E}^{(+4)}_{\ha\hb}
\nn&&
+\frac12\varepsilon_{ab}\{[(\epsilon^{-b}\theta^+_\hb)+(\bar\epsilon^{-}_\hb\bar\theta^{+b})]
{\cal D}^{++a}_\ha-[(\epsilon^{-a}\theta^+_\ha)
+(\bar\epsilon^{-}_\ha\bar\theta^{+a})]{\cal D}^{++b}_\hb\}W^{++}.
\eea

Using the component decompositions of the superfields $W^{++}$ and $V^{++a}_\ha$ from the next
section we can analyze equations \p{hWcon} and \p{hcon3} on the physical and auxiliary fields.
In particular, the field equations of the $ N=4$ supermultiplet are equivalent
to the CB superfield equations \p{DW}. All auxiliary fields  in the  superfields
$W^{++}$ and $V^{++a}_\ha$
vanish on shell or are expresed in terms of physical fields.

 In the abelian case, the on-shell
superfield constraints for the  superfield $W^{kl}$ \p{DW}
are equivalent to the following linear
harmonic-superspace equations:
\bea
&&D^{+a}_\alpha W^{++}=\bar{D}^{+}_{\ha\dot\alpha}W^{++}=0,\quad D^{++a}_\ha W^{++}=0,\\
&&
(W^{++})^\sim=-W^{++}.\lb{Wreal}
\eea
The simple HS reality constraint \p{Wreal} uses
 the harmonic  conditions
\bea
&&\frac12\varepsilon^{ijkl}\varepsilon_{ab}u^{+a}_i u^{+b}_j=
\varepsilon^{\ha\hb}\bar{u}^{+k}_\ha \bar{u}^{+l}_\hb=-(\varepsilon_{ab}u^{+a}_k u^{+b}_l)^\sim
\eea
and guarantees the self-duality condition \p{Wsd}.

The AB superfield equation
$D^{++a}_\ha W^{++}=0$ gives free equations for the physical component fields of
the  on-shell analytic superfield solution  which contains the following structures
of the $N=4$ gauge supermultiplet:
\bea
&&W^{++}\rightarrow w^{++}\sim U^{++}_{[kl]}\phi^{kl}-4i\theta^{+\alpha}_\hb\bar\theta^{+b\dot\alpha}_b
U^{b\hb}_{[kl]}\partial_{\alpha\dot\alpha}\phi^{kl}\nn
&&+[\varepsilon^{\hb\hc}\theta^{+\beta}_\hb \bar{u}^{+ k}_\hc\lambda_{k\beta}
+i\Theta^{++\alpha\beta}\bar\theta^{+b\dot\alpha}\bar{u}^{-k}_b\partial_{\alpha\dot\alpha}
\lambda_{k\beta}\nn
&&+i\Theta^{++\alpha\beta}(\partial_{\alpha\dot\alpha}A^{\dot\alpha}_{\beta}
+\partial_{\beta\dot\alpha}A^{\dot\alpha}_{\alpha})-\mbox{c.c.}]+\ldots\lb{Wsolut}
\eea
where $\phi^{[kl]}$ is the real 6-component scalar field, $A_m$ is the abelian gauge field, and
$\lambda_{k\beta}$ is the spinor field. Here we use the notation for harmonic and Grassmann
polynomials from appendices. Similar equations and solution for the abelian real
superfield $W^{++}$ were considered earlier in \cite{HH,HW2,Howe97}.

 We note that  the AB constraint equations of this section  are not derived from the action principle
, the harmonic $N=4$ transform \p{nablaon2} connects the central and analytic bases
on the mass shell.

\setcounter{equation}0
\section{\lb{3}  Action of  supergauge model in the harmonic $N=4$ superspace  and
component representation of  equations of motion}
\subsection{Superfield action with $\theta$ depending terms}

 We consider an alternative harmonic formalism of the $N=4$ gauge theory off mass shell and
 construct a gauge invariant and $SU(4)$ invariant action ($A$-model) including the independent
analytic  superfield $W^{++}$ and the prepotential
$V^{++a}_\ha$
\bea
A(W,V)\sim\frac1{g^2}\int d^{(-8)}\zeta du\mbox{Tr}\{W^{++}[(\theta^{+\ha}\theta^{+\hb})\hat{F}^{(+4)}_{\ha\hb}
+(\bar\theta^{+a}\bar\theta^{+b})F^{(+4)}_{ab}+\frac12W^{++}(\Theta^{+4}+\bar\Theta^{+4})]\}\lb{Ainv}
\eea
where $g$ is a coupling constant. The action $A$ is also invariant under the Poincare group transformations and
the scale transformations.
 Each term of $A$ explicitly depends
on the Grassmann coordinates
$\theta^+_\ha,~\bar\theta^{+a}$ , so the $N=4$ supersymmetry is broken.

Unconstrained analytic $SU(n)$ superfields satisfy the conditions
\be
[(W^{++})^A_B]^\sim=-(W^{++})_A^B,\quad [(V^{++a}_\hb)^A_B]^\sim=(V^{++b}_\ha)_A^B.
\ee

 Varying the action in the superfield $V^{++a}_\ha$ gives us the equation
\bea
&&II)\quad\mbox{dim.}0:\quad[\varepsilon_{ab}(\theta^{+\ha}\theta^{+\hb})
+\varepsilon^{\ha\hb} (\bar\theta^+_a\bar\theta^+_b)]{\cal D}^{++b}_\hb W^{++}=0\lb{EA2}
\eea
which is proportional to the covariant equation \p{hWcon} and contains the $\theta$ depending
multiplier in the brackets.

Varying $A$ in $W^{++}$ we obtain the  equation
\bea
&&II)\quad\mbox{dim.}-1:\quad(\theta^{+\ha}\theta^{+\hb})\hat{F}^{(+4)}_{\ha\hb}
+(\bar\theta^{+a}\bar\theta^{+b})F^{(+4)}_{ab}+W^{++}(\Theta^{+4}+\bar\Theta^{+4})\nn
&&=(\theta^{+\ha}\theta^{+\hb})\hat{E}^{(+4)}_{\ha\hb}
+(\bar\theta^{+a}\bar\theta^{+b})E^{(+4)}_{ab}=0\lb{EA1}
\eea
which is a  combination of two equations \p{hcon3} with nilpotent $\theta$ coefficients.

Thus, these $A$-induced equations II are not equivalent but compatible with equations I
\p{hWcon},\p{hcon3}. The component equations for physical fields are equivalent in both cases.

\subsection{Component analysis of the $A$-model}

We analyze the component decomposition of the imaginary analytic abelian gauge parameter
\bea
&&\lambda(\zeta,u)=ia(x_\A)+U^k_la^l_k(x_\A)+\theta^{+\alpha}_\hc u^{-\hc}_k\beta^k_{\alpha}(x_\A)
+\bar\theta^{+c\dot\alpha}\bar{u}^{-k}_c\bar\beta_{k\dot\alpha}(x_\A)\nn
&&+(\theta^+_\hb\theta^+_\hc)U^{--(\hb\hc)}_{(kl)}d^{(kl)}(x_\A)-(\bar\theta^{+b}\bar\theta^{+c})\tilde{U}^{(kl)}_{(bc)}
\bar{d}_{(kl)}(x_\A)
+i\theta^{+\alpha}_\hc\bar\theta^{+c\dot\alpha}U^{--\hc k}_{c l}b^{l}_{k\alpha\dot\alpha}(x_\A)\nn
&&+\Theta^{++\alpha\beta}U^{--}_{[kl]}c^{[kl]}_{(\alpha\beta)}(x_\A)
-\bar\Theta^{++\dot\alpha\dot\beta}U^{--}_{[kl]}\bar{c}^{[kl]}_{(\dot\alpha\dot\beta)}(x_\A)+\ldots
\eea
where all higher Grassmann and harmonic terms are omitted and the notation from appendices
is used. It is not difficult to consider
the component decomposition of the harmonic derivative term $D^{++a}_\ha \lambda$ \p{deltaV} and construct
 the $WZ$-type gauge condition for  the prepotential
\bea
(V^{++a}_\ha)_{WZ}=v^{++a}_\ha+{\cal V}^{++a}_\ha
\eea
where $v^{++a}_\ha$ contains
 the standard $ N=4$ supermultiplet $\phi^{[kl]},~ A_m,~\lambda_{k\alpha},~\bar\lambda^k_{\dot\alpha}$
\bea
&&v^{++a}_\ha=-2\theta^{+\alpha}_\ha\bar\theta^{+a\dot\alpha}A_{\alpha\dot\alpha}
+\phi^{[kl]}[(\theta^{+}_\ha\theta^+_{\hc})U^{a\hc}_{[kl]}
+(\bar\theta^{+a}\bar\theta^{+c})U_{c\ha[kl]}]\nn
&&+(\theta^{+}_\ha\theta^+_{\hc})\bar\theta^{+a\dot\alpha}u^{-\hc}_k
\bar\lambda^k_{\dot\alpha}
-(\bar\theta^{+a}\bar\theta^{+c})\theta^{+\alpha}_\ha \bar{u}^{-k}_c\lambda_{k\alpha},\lb{standard}
\eea
and ${\cal V}^{++a}_\ha$ includes an infinite number of
additional bosonic and fermionic component fields.
Note that the  condition $(v^{++a}_\ha)^\sim =v^{++a}_\ha$ gives us the reality of the vector field $A_m$ and the self-duality of the scalar field $\bar\phi_{[kl]}=
\frac12\varepsilon_{klij}\phi^{[ij]}$
using the self-duality of the neutral
irreducible harmonic $U^{a\ha}_{[ij]}$ \p{usdneu}
\be
\phi^{[kl]}[(\theta^{+}_\ha\theta^+_{\hc})U^{a\hc}_{[kl]}
+(\bar\theta^{+a}\bar\theta^{+c})U_{c\ha[kl]}]=\phi^{[kl]}(\theta^{+}_\ha\theta^+_{\hc})
U^{a\hc}_{[kl]}
+\bar\phi_{[kl]}(\bar\theta^{+a}\bar\theta^{+c})\tilde{U}_{\ha c}^{[kl]}.
\ee

We calculate  the standard part of the analytic superfield strength
$\hat{F}^{+4}_{\ha\hb}$
\bea
&&\hat{f}^{+4}_{\ha\hb}=\varepsilon_{ab}(D^{++a}_\ha v^{++b}_\hb
+D^{++a}_\hb v^{++b}_\ha)\nn
&&=2i\varepsilon_{ab}\varepsilon^{\beta\alpha}(\theta^{+}_\hb\theta^{+}_\ha)
\bar\theta^{+b\dot\beta}\bar\theta^{+a\dot\alpha}
(\partial_{\beta\dot\beta}A_{\alpha\dot\alpha}+
\partial_{\beta\dot\alpha}A_{\alpha\dot\beta})\nn
&&+2\varepsilon_{ab}(\theta^{+}_\ha\theta^+_{\hb})u^{+a}_k u^{+b}_l\phi^{[kl]}
-2i\varepsilon_{ab}(\theta^{+}_\hb\theta^{+}_\ha)
\theta^{+\beta}_{\hc}\bar\theta^{+b\dot\beta}
u^{+a}_k u^{-\hc}_l\partial_{\beta\dot\beta}\phi^{[kl]}\nn
&&
+2i\varepsilon_{ab}(\theta^{+\beta}_\hb u^{+k}_\ha+\theta^{+\beta}_\ha
u^{+k}_\hb)\bar\theta^{+b\dot\beta}\bar\theta^{+a}_{\dot\alpha}
\bar\theta^{+c\dot\alpha}u^{-l}_c\partial_{\beta\dot\beta}\bar\phi_{[kl]}
\nn
&&+2\varepsilon_{ab}\theta^{+\rho}_\ha\theta^+_{\rho\hb}\bar\theta^{+a\dot\alpha}
 u^{+b}_k\bar\lambda^k_{\dot\alpha}
-2i\varepsilon_{ab}(\theta^{+}_\hb\theta^{+}_\ha)\theta^{+\beta}_{\hc}
\bar\theta^{+b\dot\beta}\bar\theta^{+a\dot\alpha}u^{-\hc}_k
\partial_{\beta\dot\beta}\bar\lambda^k_{\dot\alpha}
\nn
&&
+2i\varepsilon_{ba}(\theta^{+}_\ha\theta^{+}_\hb)\bar\theta^{+b\dot\beta}
\bar\theta^{+a\dot\rho}\bar\theta^{+c}_{\dot\rho}
u^{-l}_c\partial_{\beta\dot\beta}\lambda^\beta_{l}.
\eea
The conjugated quantity $f^{(+4)ab}$ includes
$\bar\phi_{[ij]}=\frac12\varepsilon_{ijkl}\phi^{[kl]}, ~~\bar\lambda^k_{\dot\alpha}$
and the conjugated bispinor field-strength $F_{\alpha\beta}(A)$.

The prepotential contains  additional   terms with the lowest   vector and tensor dimension-1
fields having pairs of $SU(4)$ indices
\bea
&&
\theta^{+\alpha}_\ha\bar\theta^{+c\dot\alpha} U^{ak}_{cl}A^l_{k\alpha\dot\alpha}
+\theta^{+\alpha}_\hc\bar\theta^{+a\dot\alpha}\tilde{U}^{\hc k}_{\ha l}\bar{A}^l_{k\alpha\dot\alpha}\nn
&&+\Theta^{++\alpha\beta}[U^{a}_{\ha (kl)}B^{(kl)}_{(\alpha\beta)}+\tilde{U}^{a(kl)}_{\ha }B_{(kl)(\alpha\beta)}]
+\bar\Theta^{++\dot\alpha\dot\beta}[\tilde{U}^{a(kl)}_{\ha }\bar{B}_{(kl)(\dot\alpha\dot\beta)}
+U^{a}_{\ha (kl)}\bar{B}^{(kl)}_{(\dot\alpha\dot\beta)}]\lb{1add}
\eea
and additional (dimension-3/2) fermionic fields
\bea
&&\Theta^{(+3)\alpha}_\ha \bar{u}^{-ak}\xi_{k\alpha}
+\Theta^{++}_{\alpha\beta}\bar\theta^{+a\dot\alpha}u^-_{\ha k}\rho^{k(\alpha\beta)}_{\dot\alpha}
+\mbox{c.c.}
\eea

We study the lowest terms with the  (dimension-2) auxiliary fields
\bea
&&\Theta^{(+3)\alpha}_\ha\bar\theta^{+a\dot\alpha}U^{--}_{[kl]}H^{[kl]}_{\alpha\dot\alpha}
+\Theta^{(+3)\alpha}_\ha\bar\theta^{+c\dot\alpha} U^{--a(kl)}_{c}H_{(kl)\alpha\dot\alpha}\nn
&&+\Theta^{(+3)\alpha}_\hc\bar\theta^{+c\dot\alpha}[\delta_c^a U^{--\hc}_{\ha(kl)}H^{(kl)}_{\alpha\dot\alpha}+
\delta^\hc_\ha U^{--a(kl)}_{c}H_{(kl)\alpha\dot\alpha}]\nn
&&+(\theta^+_\ha\theta^+_\hc)(\bar\theta^{+a}\bar\theta^{+c})U^{--\hc k}_{cl}H^l_k
+\Theta^{++}_{\alpha\beta}(\bar\theta^{+a}\bar\theta^{+c})U^{--k}_{c\ha l}H^{l(\alpha\beta)}_{k}\nn
&&
+\Theta^{++}_{\alpha\beta}\bar\Theta^{++}_{\dot\alpha\dot\beta}U^{--a k}_{\ha l}
H^{l(\alpha\beta)(\dot\alpha\dot\beta)}_{k}
+\mbox{c.c.}
\eea

The off-shell prepotential includes also an infinite number of auxiliary fields of different
dimensions with more than two $SU(4)$ indices, for instance, the nontrivial dimensionless term
\bea
&&A^{(ij)}_{[kl]}U^{++a[kl]}_{\ha(ij)}+\mbox{c.c.}
\eea

The off-shell decomposition of the independent $\sim$-imaginary abelian superfield strength
 contains independent component fields
\bea
&&W^{++}=U^{++}_{[kl]}F^{[kl]}+\varepsilon^{\hb\hc}\theta^{+\beta}_\hb \bar{u}^{+ k}_\hc\Lambda_{k\beta}
-\varepsilon_{bc}\bar\theta^{+b\dot\alpha}u^{+c}_k\bar\Lambda^k_{\dot\alpha}
+i\Theta^{++}_{\alpha\beta}F^{\alpha\beta}+i\bar\Theta^{++}_{\dot\alpha\dot\beta}\bar{F}^{\dot\alpha\dot\beta}\nn
&&+\theta^{+\alpha}_\hb\bar\theta^{+\dot\alpha}_b[U^{b\hb}_{[kl]}
W^{[kl]}_{\alpha\dot\alpha}+U^{b\hb}_{(kl)}
W^{(kl)}_{\alpha\dot\alpha}-\tilde{U}^{b\hb(kl)}
\bar{W}_{(kl)\alpha\dot\alpha}]+(\theta^+_\hb\theta^{+\hc})\tilde{U}^{k\hb}_{\hc l}V^l_k
-(\bar\theta^{+b}\bar\theta^{+}_c)U^{ck}_{bl}\bar{V}^l_{k}\nn
&&+\Theta^{++}_{\alpha\beta}\bar\theta^{+b\dot\alpha}\bar{u}^{-k}_b R^{(\alpha\beta)}_{k\dot\alpha}
-\theta^{+\alpha}_\hb\bar\Theta^{++}_{\dot\alpha\dot\beta}u^{-\hb}_k
\bar{R}^{k(\dot\alpha\dot\beta)}_{\alpha}+
\Theta^{(+3)\alpha}_\hb u^{-\hb}_kP^k_\alpha+\bar\Theta^{(+3)b\dot\alpha}\bar{u}_b^{-k}\bar{P}_{k\dot\alpha}
\nn
&&+(\Theta^{+4}+\bar\Theta^{+4})U^{--}_{[kl]}T^{[kl]}+\Theta^{++}_{\alpha\beta}(\bar\theta^{+b}\bar\theta^{+c})\tilde{U}_{bc}^{--(kl)}
T^{(\alpha\beta)}_{(kl)}\nn
&&+\Theta^{(+3)\alpha}_\hb\bar\theta^{+b\dot\alpha}U^{--\hb k}_{b l}B^l_{k\alpha\dot\alpha}
-\theta^{+\alpha}_\hb\bar\Theta^{(+3)b\dot\alpha}U^{--\hb k}_{b l}\bar{B}^l_{k\alpha\dot\alpha}
+\ldots\lb{Wcomp}
\eea
where higher Grassmann and harmonic   terms are omitted. The fields $F^{[kl]},~W^{[kl]}_{\alpha\dot\alpha}$ and
$T^{[kl]}$ are self-dual by construction.

We stress that the nilpotency of the Grassmann coordinates yields the cancellation of some higher component terms
in the combinations $W^{++}(\theta^{+\ha}\theta^{+\hb})\hat{F}^{(+4)}_{\ha\hb}, ~~\Theta^{+4}(W^{++})^2,\ldots$ from
the action $A$.

The abelian version of eq. \p{EA2} gives us  algebraic  restrictions for the
infinite supermultiplet \p{Wcomp}
\bea
&&V^k_l=0,\quad T^{[kl]}=0,\quad P^k_\alpha=0,\ldots
\eea
 and  differential constraints
\bea
&&\partial^{\dot\alpha\beta}F_{\alpha\beta}=0,\quad \partial^{\dot\alpha\beta}\Lambda_{k\beta}=0,\quad
W^{[kl]}_{\alpha\dot\alpha}=-4i\partial_{\alpha\dot\beta}F^{[kl]},\nn
&&\partial^{\dot\alpha\alpha}W^{[kl]}_{\alpha\dot\alpha}=0,\quad
R_{k(\alpha\beta)\dot\alpha}=i\partial_{(\alpha\dot\alpha}\Lambda_{k\beta)},~\ldots
\eea
All auxiliary fields in $W^{++}$ vanish or are expressed via   the basic fields
$F_{\alpha\beta},~F^{[kl]},~\Lambda_{k\beta}$
on the mass shell, and these basic fields form the $N=4$ multiplet of auxiliary field-strengths and
satisfy the free equations of motion.

 In the abelian case, eq. \p{EA1} yields  relations between the physical and auxiliary
 field-strengths and also the restrictions  for  auxiliary fields
  in $V^{++a}_\ha$ and $W^{++}$
\bea
&&F^{[kl]}\sim\phi^{[kl]},\quad F_{\alpha\beta}\sim \partial_{\alpha\dot\alpha} A^{\dot\alpha}_{\beta}+
\partial_{\beta\dot\alpha} A^{\dot\alpha}_{\alpha},\\
&&A^{(ij)}_{[kl]}=0,\quad A^l_{k\alpha\dot\alpha}=0,\quad B^{(kl)}_{(\alpha\beta)}=0,\nn
&&
\rho^k_{(\alpha\beta)\dot\alpha}=0,\ldots
\eea
Compairing two equations we obtain the free equations for the physical fields
$\phi^{[kl]},~~A_m$ and $\lambda_{k\alpha}$ in the abelian case,
and all auxiliary fields vanish or are expressed via the physical fields on the mass shell.
The non-abelian nonlinear equations \p{EA2} and \p{EA1} preserve gauge covariance of the algebraic and differential
constraints. Excluding the auxiliary fields we obtain  the standard $N=4$ Yang-Mills equations
for the physical component fields.

The proof of consistency of the abelian component action in the
$A$-model and its gauge invariance are sufficient for the proof of consistency of the
corresponding nonabelian component action. The possible superfield quantization
of the $A$-model will be discussed elsewhere.

\setcounter{equation}0
\section{Nonlinear  interactions in the abelian $N=4$ gauge theory}

We consider the bilinear component electromagnetic terms from the abelian version $A_0$
of the action  \p{Ainv}
\bea
&&\frac1{64}[F^{\alpha\beta}F_{\alpha\beta}+4F^{\alpha\beta}(\partial_{\alpha\dot\beta}A^{\dot\beta}_{\beta}+
\partial_{\beta\dot\beta}A^{\dot\beta}_{\alpha})+\mbox{c.c}],\\
&&F^2=F^{\alpha\beta}F_{\alpha\beta},\quad \bar{F}^2=\bar{F}^{\dot\alpha\dot\beta}\bar{F}_{\dot\alpha\dot\beta}.
\eea
Excluding auxiliary fields $F_{\alpha\beta}$ and $\bar{F}_{\dot\alpha\dot\beta}$ we obtain the standard electromagnetic Lagrangian
 $-\frac14(\partial_mA_n-\partial_nA_m)^2$. Excluding of other auxiliary fields in the component version
of the action $A_0$ yields the standard abelian action of the physical $N=4$ fields.

  The $N=4$
supersymmetric fourth-order interaction of the abelian harmonic superfield
\be
S_4=N_1f^2\int d^{(-8)}\zeta du (W^{++})^4
\ee
contains the coupling constant $f$ of a dimension $-2$ and some normalization factor $N_1$ .
This term
describes the simplest effective  interaction  by analogy
with the nonlinear abelian $N=3$ gauge model \cite{IZ}. In particular, this superfield term
yields the component interaction
\be
L_4\sim f^2F^2\bar{F}^2
\ee
and similar fourth-order interactions for other auxiliary component fields.
The  simplest $N=4$ effective action $A_0+S_4$ describes nonpolynomial interactions
of physical fields, if we exclude auxiliary fields. We see that auxiliary fields play
the important role in the study of effective interactions of
 $N=4$ supermultplets.

By  analogy with \cite{IZ}, we can construct the $U(1)$ neutral analytic superfield ${\cal A}$
of a dimension 8
\bea
{\cal A}=N_2D^{+4}\bar{D}^{+4}[(D^{--\ha}_a D^{--a}_\ha)^4(W^{++})^4].
\eea
We choose the constant $N_2$ by the normalization condition of the component decomposition
\be
{\cal A}=F^2\bar{F}^2+\ldots
\ee
where all other terms are omitted. The nonlinear interaction of the superfield $W^{++}$
is defined by an arbitrary
dimensionless function of this analytic superfield
\bea
&&S(W)=f^2\int d^{(-8)}\zeta du (W^{++})^4E(f^4{\cal A})\nn
&&=f^2\int d^{(-8)}\zeta du (W^{++})^4
[N_1+e_1f^4{\cal A}+e_2f^8{\cal A}^2+\ldots]
\eea
where $e_1,~e_2,...$ are some constants.

We stress that the superfield interaction $S(W)$ generates  consistent equations in
combination with the bilinear
abelian action $A_0$ \p{Ainv}.
Varying $V^{++a}_\ha$ gives the linear eq. \p{EA2}, and varying  $W^{++}$ leads
 to a nonlinear generalization of eq. \p{EA1}.
Excluding the auxiliary fields of $W^{++}$ from the component decomposition of these
superfield interactions
we obtain nonlinear effective interactions of the physical $N=4$ supermultiplet.
These interactions describe possible quantum corrections to the $N=4$ classical action.

\setcounter{equation}0
\section{\lb{5}Inconsistencies of the manifestly covariant model in  harmonic  superspace}

We analyze the simplest superfield invariant of the gauge and superconformal groups
using the gauge prepotential $V^{++a}_\ha$
\bea
S=\int d^{(-8)}\zeta du\mbox{Tr}[F^{(+4)ab}F^{(+4)}_{ab}+\hat{F}^{(+4)}_{\ha\hb}\hat{F}^{(+4)\ha\hb}]
\lb{Sinv}
\eea
as an action of some gauge   theory ($S$-model)\footnote{
Similar invariant actions were considered in the $USp(4)$ harmonic superspace formalism
\cite{BLS}.}.

Varying the prepotential $V^{++a}_\hb$ we obtain the equation of motion of the $S$-model
\bea
&&{\cal D}^{++a\ha}\hat{F}^{+4}_{\hb\ha}+{\cal D}^{++}_{b\hb}F^{+4ab}=0.
\eea

The abelian version of this equation has the form
\bea
&&\varepsilon_{ce}D^{++a\ha}(D^{++e}_\ha V^{++c}_\hb-D^{++c}_\hb V^{++e}_\ha)
+\varepsilon^{\hc\he}D^{++}_{b\hb}(D^{++a}_\he V^{++b}_\hc-D^{++b}_\hc V^{++a}_\he)\nn
&&=2D^{++}_{c\hb}D^{++a\hc} V^{++c}_\hc
-D^{++c}_\hc D^{++\hc}_c  V^{++a}_\hb=0.
\eea

It is easy to analyze  the component decomposition of the superfield abelian action and
equation of motion. The standard component fields from $v^{++a}_\ha$ \p{standard} have the
reasonable interactions. The additional dimension-1 fields \p{1add} satisfy dynamical
differential equations of motion, however, we do not find additional gauge symmetries for
interactions of unusual vector fields $A^k_{l\alpha\dot\alpha}$
and tensor fields $B^{(kl)}_{(\alpha\beta)}$ in this $S$-model. We know that the gauge invariance
$\delta A^k_{l\alpha\dot\alpha}=\partial_{\alpha\dot\alpha}\Lambda^k_l$ is necessary for a
consistent description of the free massless vector field with a positive energy, and similar
gauge invariance is important for
tensor fields. Thus, the component decomposition of the
covariant superfield action is inconsistent  at the bilinear level for these unusual fields.
We conclude that the manifestly covariant superfield action $S$  describes the inconsistent
gauge field interaction of the standard $N=4$ supermultiplet with additional bosonic and fermionic
fields.

\setcounter{equation}0
\section{Conclusions}

We reformulate the superfield constraints of the $N=4$ gauge theory \cite{So} in the formalism
of the ${\cal U}_8$ harmonic superspace ${\cal H}(4+8|8)$. In the nonabelian case,
the harmonic-superspace equations connect the independent dimension-1 harmonic superfield $W^{++}$
and the dimensionless gauge prepotential $V^{++a}_\ha$. These superfield equations I explicitly
depend on the Grassmann coordinates, although they are covariant with respect to deformed $N=4$
supersymmetry transformations of prepotentials.

We use the unconstrained superfields $W^{++}$ and  $V^{++a}_\ha$ in the nonabelian superfield
action $A$. This action explicitly depends  on the Grassmann coordinates, although the $SU(4)$
 automorphism symmetry is preserved. At the field-component level, the action $A$ describes
 interactions of two infinite-dimensional $N=4$ supermultiplets. The $A$-model equations of
motion II are compatible with the harmonic-superspace constraint equations.
On mass shell, all auxiliary fields vanish or are expressed
in terms of the physical $N=4$ fields $\phi^{[kl]},~A_m$ and $\lambda_{k\alpha}$.

Possible quantum corrections in the $N=4$ theory are described by the nonlinear
effective interaction of the superfield $W^{++}$.

\section*{Acknowledgements}
This work is partially supported  by the RFBR grants Nr.12-02-00517,
Nr.13-02-91330, Nr.13-02-90430, the grant DFG LE 838/12-1 and  grant of the Heisenberg-Landau
program. The author is grateful to E.A. Ivanov for discussions.

\def\theequation{A.\arabic{equation}}
\setcounter{equation}0
\section*{\lb{I} Appendix A. $SU(4)/SU(2)\times SU(2)\times U(1)$ harmonic coset space}
\subsection*{A.1 Harmonic variables and harmonic derivatives}
We use the
 harmonics  parameterizing the
 8 dimensional coset space ${\cal U}_8=G/H$
\bea
u^{+a}_k
~,\quad u^{-\hat{a}}_k
\eea
where $G=SU(4)$,
$H=SU_L(2)\times SU_R(2)\times U(1)$, $k=1,2,3,4$ is the spinor index  of $SU(4)$, $a=1,2$ describe
the $SU_L(2)$ doublet and $\hat{a}=\hat{1},\hat{2}$ corresponds to the $SU_R(2)$ doublet,
and $\pm$ are charges of $U(1)$. They form an $SU(4)$ matrix and are
covariant under the independent $G$ and $H$ transformations.

The Hermitian conjugation of the harmonic matrix $SU(4)$ gives us  the conjugated harmonics
\bea
&&\overline{u^{+a}_k}=\bar{u}^{-k}_a~,\quad\overline{u^{-\ha}_k}=\bar{u}^{+k}_{\hat{a}}.~\lb{I1}
\eea

The harmonics satisfy the following basic relations \cite{IKNO} :
\bea
&&u^{+a}_i
\bar{u}^{-i}_b=\delta^a_b~,\quad u^{+a}_i \bar{u}^{+i}_\hb=0,\lb{A4}\\
&&u^{-\hat{a}}_i \bar{u}^{+i}_{\hat{b}}=\delta^\ha_\hb,\quad
u^{-\hat{a}}_i \bar{u}^{-i}_b=0~,\lb{A5}\\
&&u^{+a}_i \bar{u}^{-k}_a+u^{-\ha}_i \bar{u}^{+k}_\ha=\delta^k_i,\lb{A6}\\
&& \varepsilon^{ijkl}u^{+a}_i u^{+b}_j u^{-\ha}_k
u^{-\hb}_l=\varepsilon^{ab}\varepsilon^{\ha\hb}
~.\lb{A7}
\eea

The special $\sim$-conjugation  for these
harmonics and other quantities  preserves $U(1)$ charge and changes indices of two $SU(2)$ subgroups
\bea
&&(u^{+a}_i)^\sim=-\bar{u}^{+i}_{\hat{a}}~,\quad
(u^{-\hat{a}}_i)^\sim=-\bar{u}^{-i}_a~,\quad (\varepsilon^{ab})=\varepsilon_{\hb\ha},\nn
&&
(\bar{u}^{-i}_a)^\sim=-u^{-\hat{a}}_i~,\quad (\bar{u}^{+i}_{\hat{a}})^\sim=
-u^{+a}_i,\lb{A8}
\eea
it is consistent with the basic harmonic relations.
Note that
our convention of $\sim$-conjugation uses an additional sign $-$ in comparison with \cite{IKNO}.

The $U(1)$ neutral $SU(4)$-invariant harmonic derivatives
\bea
&&\partial^a_b=u^{+a}_i\partial^{-i}_b-\bar{u}^{-i}_b\bar\partial^{+a}_i
-\frac12\delta^a_b(u^{+f}_i\partial^{-i}_f-\bar{u}^{-i}_f
\bar\partial^{+f}_i),\\
&&
\hat{\partial}^\ha_\hb=u^{-\ha}_i\partial^{+i}_\hb-\bar{u}^{+i}_\hb
\bar\partial^{-\ha}_i
-\frac12\delta^\ha_\hb (u^{-\hf}_i\partial^{+i}_\hf
-\bar{u}^{+i}_\hf\bar\partial^{-\hf}_i),\\
&&\partial^0=u^{+f}_i\partial^{-i}_f-\bar{u}^{-i}_f
\bar\partial^{+f}_i+\bar{u}^{+i}_\hf\bar\partial^{-\hf}_i
-u^{-\hf}_i\partial^{+i}_\hf
\eea
contain partial harmonic derivatives
\bea
&&\partial^{-i}_b=\frac{\partial}{\partial u^{+b}_i},\quad \partial^{+i}_\hb
=\frac{\partial}{\partial u^{-\hb}_i},\quad
\bar\partial^{-\hb}_i=\frac{\partial}{\partial \bar{u}^{+i}_\hb},\quad
\bar\partial^{+b}_i=\frac{\partial}{\partial \bar{u}^{-i}_b}.
\eea
These neutral harmonic derivatives satisfy the $SU_L(2)\times SU_R(2)\times U(1)$
Lie algebra.

The eight charged  coset harmonic derivatives
\bea
&&\partial^{++a}_\hb=u^{+a}_i\partial^{+i}_\hb-\bar{u}^{+i}_\hb
\bar\partial^{+a}_i,\\
&&\partial^{--\ha}_b=u^{-\ha}_i\partial^{-i}_b-\bar{u}^{-i}_b
\bar\partial^{-\ha}_i.
\eea
satisfy the $SU(4)$ Lie algebra
\bea
&&[\partial^{++a}_\ha,\partial^{--\hb}_b]=\partial^a_b\delta_\ha^\hb -\hat{\partial}^\hb_\ha\delta^a_b
+\frac12\delta^a_b\delta^\hb_\ha\partial^0~,\lb{I8}\\
&&[\partial^b_c,\partial^{++a}_\ha]=\delta^a_c\partial^{++b}_\ha-\frac12\delta^b_c\partial^{++a}_\ha,\quad
[\partial^\hb_\hc,\partial^{++a}_\ha]=-\delta^\hb_\ha\partial^{++a}_\hc+\frac12\delta^\hb_\hc\partial^{++a}_\ha,\nn
&&[\partial^b_c,\partial^{--\ha}_a]=-\delta^b_a\partial^{--\ha}_c+\frac12\delta^b_c\partial^{--\ha}_a,\quad
[\partial^\hb_\hc,\partial^{--\ha}_a]=\delta^\ha_\hc\partial^{--\hb}_\hc-\frac12\delta^\hb_\hc\partial^{--\ha}_a,\nn
&&\quad [\partial_0,\partial^{++a}_\ha]=2\partial^{++a}_\ha,\quad
\quad [\partial_0,\partial^{--\ha}_a]=-2\partial^{--\ha}_a.
\eea

The special Hermitian conjugation is defined for the harmonic derivatives
\bea
&&(\partial^{+i}_\hb)^\dagger=\bar\partial^{+b}_i,\quad (\partial^{-i}_b)^\dagger=\bar\partial^{-\hb}_i,\\
&&(\partial^a_b)^\dagger=\hat{\partial}^\hb_\ha,\quad(\partial^0)^\dagger=-\partial^0,\quad
(\partial^{++a}_\hb)^\dagger=\partial^{++b}_\ha,\quad (\partial^{--\ha}_b)^\dagger=\partial^{--\hb}_a,\\
&&(\partial^{++a}_\hb f)^\sim
=[\partial^{++a}_\hb, f]^\dagger=-\partial^{++b}_\ha\tilde{f},\quad
(\partial^a_b f)^\sim
=[\partial^a_b, f]^\dagger=-\hat{\partial}^\hb_\ha\tilde{f}
\eea
where $f$ and $\tilde{f}$ are the $\sim$-conjugated harmonic functions.

These $\sim$-conjugation rules are compatible with the following formulas:
\bea
&&\partial^a_b u^{+c}_i=u^{+a}_i\delta_b^c-\frac12\delta^a_bu^{+c}_i,\quad
\partial^a_b \bar{u}^{-i}_c=-\bar{u}^{-i}_b\delta_c^a+
\frac12\bar{u}^{-i}_c\delta^a_b,\nn
&&\hat{\partial}^\ha_\hb u^{-\hc}_i=u^{-\ha}_i\delta_\hb^\hc
-\frac12\delta^\ha_\hb u^{+\hc}_i,\quad
\hat{\partial}^\ha_\hb \bar{u}^{+i}_\hc=-\bar{u}^{+i}_\hb\delta_\hc^\ha
+\frac12\bar{u}^{+i}_\hc\delta^\ha_\hb,\\
&&\partial^0u^{+c}_i=u^{+c}_i,\quad \partial^0\bar{u}^{-i}_c=
-\bar{u}^{-i}_c,\quad
\partial^0u^{-\hc}_i=-u^{-\hc}_i,\quad \partial^0\bar{u}^{+i}_\hc=
\bar{u}^{+i}_\hc,\nn
&&\partial^{++a}_\hb \bar{u}^{-i}_c=-\delta^a_c \bar{u}^{+i}_\hb,\quad
\partial^{++a}_\hb u^{-\hc}_i=\delta^\hc_\hb u^{+a}_i,\nn
&&
\partial^{--\ha}_bu^{+c}_i=\delta^c_bu^{-\ha}_i,~~
\partial^{--\ha}_b \bar{u}^{+i}_\hc=
-\delta^\ha_\hc \bar{u}^{-i}_b.~~~
\eea

\subsection*{A.2 Irreducible harmonic combinations}

We study irreducible in  $SU(4)$ and $SU_L(2)\times SU_R(2)$ group indices combinations
of the harmonic coordinates with diferent $U(1)$ charge $q$
\begin{itemize}
{\bf \item ~~$q=0$ harmonics}\\

 We consider the bilinear traceless neutral combinations of harmonics
\bea
&&U^k_l=u^{+b}_l \bar{u}^{-k}_b-\bar{u}^{+k}_\hb u^{-\hb}_l=-(U^l_k)^\sim,\quad U^k_lU^l_j=\delta^k_j.
\eea

The completely traceless Hermitian neutral combination with four $SU(4)$ indices
\bea
&&U^{ik}_{jl}=\frac14U^k_lU^i_j+\frac1{60}\delta^k_l\delta^i_j-\frac1{15}\delta^k_j\delta^i_l
=(U_{ik}^{jl})^\sim
\eea
has the combined symmetry $U^{ik}_{jl}=U^{ki}_{lj}$.

The simplest bilinear combinations with $SU_L(2)\times SU_R(2)$ indices have the form
\bea
&&U^{a\ha}_{[ij]}=\frac12u^{+a}_i u^{-\ha}_j-\frac12u^{+a}_j u^{-\ha}_i,\quad
U^{a\ha}_{(ij)}=\frac12u^{+a}_i u^{-\ha}_j+\frac12u^{+a}_j u^{-\ha}_i,\\
&&\tilde{U}^{[ij]}_{a\ha}=(U^{a\ha}_{[ij]})^\sim,\quad \tilde{U}^{(ij)}_{a\ha}=(U^{a\ha}_{(ij)})^\sim.
\eea

We consider the important $U(1)$ neutral self-duality relation connecting conjugated harmonics
\bea
&&U^{a\ha}_{[ij]}=\frac12\varepsilon_{ijkl}\varepsilon^{ab}\varepsilon^{\ha\hb}
\tilde{U}^{[kl]}_{b\hb}.\lb{usdneu}
\eea
Note that the determinant condition \p{A7} follows from this relation.

The double traceless bilinear neutral combinations
\bea
&&U^{ak}_{bl}=u^{+a}_l\bar{u}^{-k}_b-\frac12\delta^a_b
u^{+e}_l\bar{u}^{-k}_e,\quad
\tilde{U}^{\hb l}_{\ha k}=(U^{ak}_{bl})^\sim
\eea
satisfy the relations $U^{ak}_{bl}U_k^j=U^{aj}_{bl}=U^{aj}_kU^k_l$.
We can construct the traceless combination with four $SU(4)$ indices.
\\

{\bf \item~~$q=1$ harmonics}\\

We construct the traceless combination with three $SU(4)$ indices
\bea
&&U^{+ai}_{jl}=\frac12u^{+a}_j u^{+b}_l\bar{u}^{-i}_b
-\frac2{30}\delta^i_j u^{+a}_l-\frac7{30}\delta^i_l u^{+a}_j=U^{ik}_{lj}u_k^{+a}
\eea
and the conjugated combination
$\tilde{U}^{+jl}_{\ha i}=(U^{+al}_{jl})^\sim$.
\\

{\item\bf $q=2$ harmonics}\\

The $q=2$ charged self-duality condition
\be
\frac12\varepsilon^{klij}U^{++}_{[ij]}=\tilde{U}^{++[kl]}\lb{usdplus}
\ee
connects the corresponding   combinations of harmonics
\bea
&&U^{++}_{[ij]}=\varepsilon_{ab}u^{+a}_i u^{+b}_j,\quad \tilde{U}^{++[ij]}=
\varepsilon^{\hc\he}\bar{u}^{+i}_\hc \bar{u}^{+j}_\he=-(U^{++}_{[ij]})^\sim.\lb{uplus2}
\eea

We consider the bilinear combinations with two $SU(4)$ indices
\bea
&&U^{++(ab)}_{(ij)}=\frac12(u^{+a}_iu^{+b}_j+u^{+b}_iu^{+a}_j),\quad
\tilde{U}^{++(ij)}_{(\ha\hb)}=\frac12(\bar{u}^{+i}_\ha \bar{u}^{+j}_\hb+\bar{u}^{+i}_\hb \bar{u}^{+j}_\ha),\\
&&U^{++bj}_{\hb i}=u^{+b}_i\bar{u}^{+j}_\hb.
\eea

We calculate the harmonic derivative of neutral harmonics
\bea
&&\partial^{++a}_\ha U^k_l=-2U^{++ak}_{\ha l},\\
&&\partial^{++a}_\ha U^{b\hb}_{[ij]}=\frac12\delta^\hb_\ha \varepsilon^{ab}U^{++}_{[ij]},\\
&&\partial^{++a}_\ha U^{b\hb}_{(ij)}=\delta^\hb_\ha U^{++(ab)}_{(ij)},\quad
\partial^{++a}_\ha \tilde{U}_{b\hb}^{(ij)}=-\delta^a_b \tilde{U}^{++(ij)}_{(\ha\hb)},\\
&&\partial^{++a}_\ha U^{bk}_{cl}=-\delta^a_cu^{+b}_l\bar{u}^{+k}_\ha+\frac12\delta^b_cu^{+a}_l\bar{u}^{+k}_\hb,\nn
&&\partial^{++a}_\ha \tilde{U}_{\hb k}^{\hc l}=\delta^\hc_\ha u^{+a}_k\bar{u}^{+l}_\hb
-\frac12\delta^\hc_\hb u^{+a}_k\bar{u}^{+l}_\ha.
\eea
The  traceless combinations with four $SU(4)$ indices have the form
\bea
&&U^{++a(jl)}_{\ha (ik)}=\partial^{++a}_\ha U^{(jl)}_{(ik)},\quad
U^{++a[jl])}_{\ha [ik]}=\partial^{++a}_\ha U^{[jl]}_{[ik]},\\
&&U^{++a[jl]}_{\ha (ik)}=\frac14(U^{++aj}_{\ha i}U^l_k+U^{++aj}_{\ha k}U^l_i-
U^{++al}_{\ha i}U^j_k-U^{++al}_{\ha k}U^j_i)\nn
&&+\frac1{8}(\delta^j_kU^{++al}_{\ha i}+\delta^j_iU^{++al}_{\ha k}-\delta^l_iU^{++aj}_{\ha k}
-\delta^l_kU^{++aj}_{\ha i}).
\eea
Note that the harmonic $U^{++a[jl]}_{\ha (ik)}$
cannot be presented as a total $\partial^{++a}_\ha$ derivative of a neutral harmonic.
\\

{\bf \item~~$q=-1$ harmonics}\\

We construct the traceless combination with three $SU(4)$ indices
\bea
&&U_{bl}^{-jk}=\frac12\bar{u}^{-j}_b u^{+c}_l\bar{u}^{-k}_c
-\frac2{30}\delta^j_l \bar{u}^{-k}_b-\frac7{30}\delta^k_l \bar{u}^{-j}_b
\eea
and the conjugated combination.

{\bf\item $q=-2$ harmonics}\\

The $q=-2$ charged self-duality condition
\be
\frac12\varepsilon^{klij}U^{--}_{[ij]}=\tilde{U}^{--[kl]}
\ee
connects the corresponding  combinations of harmonics
\bea
&&U^{--}_{[ij]}=\varepsilon_{\ha\hb}u^{-\ha}_i u^{-\hb}_j,\quad \tilde{U}^{--[ij]}=
\varepsilon^{ce}\bar{u}^{-i}_c \bar{u}^{-j}_e=-(U^{--}_{[ij]})^\sim,\nn
&&
\quad
\varepsilon^{ijkl}U^{--}_{[ij]}U^{++}_{[kl]}=4.\lb{usdminus}
\eea
We use the relation
\bea
&&\partial^{++a}_\ha U^{--}_{[ij]}=2\varepsilon_{\ha\hb}U^{a\hb}_{[ij]}.
\eea

We consider the bilinear combinations with two $SU(4)$ indices and two doublet indices
\bea
&&U^{--(\ha\hb)}_{(ij)}=\frac12(u^{-\ha}_iu^{-\hb}_j+u^{-\hb}_iu^{-\ha}_j),\quad
\tilde{U}^{--(ij)}_{(ab)}=\frac12(\bar{u}^{-i}_a\bar{u}^{-j}_b+\bar{u}^{-i}_b\bar{u}^{-j}_a),\\
&&U^{--\ha j}_{a i}=u^{-\ha}_i\bar{u}^{-j}_a
\eea
satisfying the relations
\bea
&&\partial^{++a}_\ha U^{--(\hc\hb)}_{(ij)}=\frac12\delta^\hc_\ha( u^{+a}_iu^{-\hb}_j
+u^{-\hb}_iu^{+a}_j)
+\frac12\delta^\hb_\ha( u^{-\hc}_iu^{+a}_j+ u^{+a}_iu^{-\hc}_j),\\
&&\partial^{++a}_\ha U^{--\hb j}_{b i}=\delta^\hb_\ha (u^{+a}_i\bar{u}^{-j}_b
-\frac12\delta^a_bu^{+c}_i\bar{u}^{-j}_c)
-\delta^a_b (u^{-\hb}_i\bar{u}^{+j}_\ha-\frac12\delta^\hb_\ha u^{-\hc}_i\bar{u}^{+j}_\hc)\nn
&&+\frac12\delta^a_b\delta^\hb_\ha (u^{+c}_i\bar{u}^{-j}_c-u^{-\hc}_i\bar{u}^{+j}_\hc)
=\frac12\delta^a_b\delta^\hb_\ha U^j_i+\delta^\hb_\ha U^{aj}_{bi}-\delta^a_b\tilde{U}^{\hb j}_{\ha i}.
\eea

\end{itemize}

\def\theequation{B.\arabic{equation}}
\setcounter{equation}0
\section*{ Appendix B. $SU(4)/SU(2)\times SU(2)\times U(1)$ harmonic superspace}
\subsection*{B.1 Analytic basis}

We use the $N=4$ superspace $R(4|16)$
with 4 space-time and 8+8 spinor coordinates
$$z=(x^m,~\theta_k^\alpha ,~\bar\theta^{k\dot\alpha})\lb{B3b}$$
where $k=1,2,3,4$ is the $SU(4)$ index, $m=0,1,2,3$ is the vector index and $\alpha,~\dot\alpha$
are the $SL(2,C)$ spinor indices.
The supersymmetry transformations have the form
\bea
\delta x^m=-i(\epsilon_k\sigma^m\bar\theta^k)+i(\theta_k\sigma^m\bar\epsilon^k),
\quad\delta\theta^\alpha_k=\epsilon^\alpha_k,\quad\delta \bar\theta^{k\dot\alpha}=
\bar\epsilon^{k\dot\alpha}.
\eea

Spinor derivatives satisfy the relations
\bea
&&\{D^k_\alpha, D^l_\beta\}=0,\quad \{\bar{D}_{k\dot\alpha}, \bar{D}_{l\dot\beta}\}=0,\nn
&&
\quad \{D^k_\alpha, \bar{D}_{l\dot\alpha}\}=-2i\delta^k_l(\sigma^m)_{\alpha\dot\alpha}\partial_m=-2i\delta^k_l\partial_{\alpha\dot\alpha}.
\eea

We construct the analytic coordinates $\zeta=(x_\A^m,~~\theta^{+\alpha}_\ha,~~
\bar\theta^{+a\dot\alpha})$
 in the ${\cal U}_8$ harmonic superspace
${\cal H}(4+8|8)$
\bea
&&x_\A^m=x^m-i(\theta^-_a\sigma^m\bar\theta^{+a})
+i(\theta^+_\ha\sigma^m\bar\theta^{-\ha}),\\
&&\delta_\epsilon x^m_\A=2i(\theta^+_\ha\sigma^m\bar\epsilon^{-\ha})
-2i(\epsilon^-_a\sigma^m\bar\theta^{+a}),\\
&&\theta^{+\alpha}_\ha=\theta^\alpha_i \bar{u}^{+i}_\ha,\quad
\bar\theta^{+a\dot\alpha}=\bar\theta^{i\dot\alpha}u^{+a}_i=
-(\theta^{+\alpha}_\ha)^\sim.
\eea
The Grassmann-analytic superfields $\lambda(\zeta,u)$ are defined in the analytic superspace.
We add also the nonanalytic spinor coordinates in the analytic basis (AB)
\bea
&&\theta^{-\alpha}_a=\theta^\alpha_i \bar{u}^{-i}_a,\quad
\bar\theta^{-\ha\dot\alpha}=\bar\theta^{i\dot\alpha}u^{-\ha}_i=-(\theta^{-\alpha}_a)^\sim,\quad
(\theta^{-a\alpha})^\sim=\bar\theta^{-\dot\alpha}_\ha.
\eea

We consider bilinear products of spinor coordinates
\bea
&&\theta^{+\alpha}_\ha\theta^{+\beta}_\hb=\frac12\varepsilon^{\beta\alpha}
(\theta^+_\ha\theta^+_\hb)+\frac12\varepsilon_{\ha\hb}
\Theta^{++\alpha\beta},\\
&&\Theta^{++\alpha\beta}=\varepsilon^{\hb\ha}\theta^{+\alpha}_\ha
\theta^{+\beta}_\hb,\\
&&\Theta^{++\alpha\beta}(\theta^+_\ha\theta^+_\hb)=0
\eea
and analogous relations for conjugated quantities
\bea
&&\bar\theta^{+a\dot\alpha}\bar\theta^{+b\dot\beta}=\frac12\varepsilon^{\dot\alpha\dot\beta}
(\bar\theta^{+a}\bar\theta^{+b})+\frac12\varepsilon^{ab}\bar\Theta^{++\dot\alpha\dot\beta},\\
&&\bar\Theta^{++\dot\alpha\dot\beta}=(\Theta^{++\alpha\beta})^\sim
=\varepsilon_{ba}\bar\theta^{+a\dot\alpha}\bar\theta^{+b\dot\beta}.
\eea

The third degree relations read
\bea
&&\Theta^{(+3)\alpha}_\ha=(\theta^+_\ha\theta^+_\hb)\theta^{+\hb\alpha}=
\varepsilon^{\hc\hb}\theta^{+}_{\ha\beta}\theta^{+\beta}_{\hb}\theta^{+\alpha}_\hc
=\theta^{+}_{\ha\beta}\Theta^{++(\beta\alpha)},\\
&&(\theta^+_\ha\theta^+_\hb)\theta^{+\alpha}_\hc=\frac13\varepsilon_{\hc\ha}
\Theta^{(+3)\alpha}_\hb+\frac13\varepsilon_{\hc\hb}\Theta^{(+3)\alpha}_\ha,\\
&&\theta^{+\alpha}_{\ha}\Theta^{++\beta\rho}=\frac13\varepsilon^{\alpha\beta}\Theta^{(+3)\rho}_\ha
+\frac13\varepsilon^{\alpha\rho}\Theta^{(+3)\beta}_\ha,\\
&&\bar\Theta^{(+3)a\dot\alpha}=-(\Theta^{(+3)\alpha}_\ha)^\sim=(\bar\theta^{+a}\bar\theta^{+b})
\bar\theta^{+\dot\alpha}_b.
\eea
\bea
&&\theta^{+\alpha}_\ha\bar\theta^{+a\dot\alpha}\theta^{+\beta}_\hb=\frac12[\varepsilon^{\alpha\beta}
(\theta^+_\ha\theta^+_\hb)-\varepsilon_{\ha\hb}\Theta^{++\alpha\beta}]\bar\theta^{+a\dot\alpha}.
\eea
We consider the 4th degree relations
\bea
&&\Theta^{+4}=(\theta^+_\ha\theta^+_\hb)(\theta^{+\ha}\theta^{+\hb})=
\Theta^{++\alpha\beta}\Theta^{++}_{\alpha\beta},\\
&&(\theta^+_\ha\theta^+_\hb)(\theta^+_\he\theta^+_\hc)=\frac16
(\varepsilon_{\ha\he}\varepsilon_{\hb\hc}+\varepsilon_{\hb\he}
\varepsilon_{\ha\hc})\Theta^{+4},\\
&&\Theta^{(+3)\alpha}_\ha\theta^{+\beta}_\hc=
\frac12\varepsilon^{\hb\he}\varepsilon^{\beta\alpha}
(\theta^+_\ha\theta^+_\hb)(\theta^+_\he\theta^+_\hc)=\frac12\varepsilon^{\alpha\beta}\varepsilon_{\ha\hc}\Theta^{+4},\\.
&&\bar\Theta^{+4}=(\bar\theta^{+\ha}\bar\theta^{+b})(\bar\theta^{+}_a\bar\theta^{+}_b)
=\bar\Theta^{++\dot\alpha\dot\beta}\bar\Theta^{++}_{\dot\alpha\dot\beta}.
\eea

We use the  spinor derivatives in the analytic basis
\bea
&&D^{+b}_{\alpha}=\partial^{+b}_{\alpha},\quad\bar{D}^{+}_{\hb\dot\alpha}
=-\bar{\partial}^{+}_{\hb\dot\alpha},\quad \\
&&D^{-\ha}_{\alpha}=\partial^{-\ha}_{\alpha}+2i\bar\theta^{-\ha\dot\alpha}
\partial^\A_{\alpha\dot\alpha},\quad
\bar{D}^-_{b\dot\alpha}=-\bar\partial_{b\dot\alpha}^{-}-
2i\theta^{-\alpha}_b\partial^\A_{\alpha\dot\alpha},\\
&&\partial^{+b}_{\alpha}=\frac{\partial}{\partial\theta^{-\alpha}_b},\quad \partial^{-\ha}_{\alpha}
=\frac{\partial}{\partial\theta^{+\alpha}_\ha},\quad
\bar{\partial}^{+}_{\hb\dot\alpha}=\frac{\partial}{\partial\bar\theta^{-\hb\dot\alpha}},
\quad
\bar{\partial}^{-}_{b\dot\alpha}=\frac{\partial}{\partial\bar\theta^{+b\dot\alpha}},\nn
&&\partial^\A_{\alpha\dot\alpha}=(\sigma^m)_{\alpha\dot\alpha}\frac{\partial}{\partial x^m_\A}.
\eea

The CR harmonic derivative in AB
\bea
D^{++b}_\hb=\partial^{++b}_\hb+2i\theta^{+\beta}_\hb\bar\theta^{+b\dot\beta}
\partial_{\beta\dot\beta}^\A-\theta^{+\beta}_\hb\partial^{+b}_{\beta}
+\bar\theta^{+b\dot\beta}\bar\partial_{\hb\dot\beta}^{+}=(D^{++b}_\hb)^\dagger,
\eea
preserves analyticity
\bea
 [D^{++b}_\hb,D^{+a}_{\alpha}]=0,\quad  [D^{++b}_\hb,\bar{D}^{+}_{\ha\dot\alpha}]=0.
\eea

We also define the nonanalytic and neutral harmonic derivatives in AB
\bea
&&D^{--\ha}_b=\partial^{--\ha}_b-2i(\theta^-_b\sigma^m\bar\theta^{-\ha})\partial_m^A
-\theta^{-\alpha}_b\partial_{\alpha}^{-\ha} +\bar\theta^{-\ha\dot\alpha}
\bar\partial_{b\dot\alpha}^{-},\\
&&D^0=\partial^0+\theta^{+\alpha}_\ha \partial^{-\ha}_\alpha+\bar\theta^{+a\dot\alpha} \bar\partial^{-}_{a\dot\alpha}
-\theta^{-\alpha}_a \partial^{+a}_\alpha-\bar\theta^{-\ha\dot\alpha} \bar\partial^{+}_{\ha\dot\alpha},\\
&&D^a_b=\partial^a_b+\bar\theta^{+a\dot\alpha} \bar\partial^{-}_{b\dot\alpha}-\frac12\delta^a_b
\bar\theta^{+c\dot\alpha} \bar\partial^{-}_{c\dot\alpha}
-\theta^{-\alpha}_b \partial^{+a}_\alpha+\frac12\delta^a_b\theta^{-\alpha}_c \partial^{+c}_\alpha,\\
&&\hat{D}^\ha_\hb=\partial^\ha_\hb-\theta^{+\alpha}_\hb \partial^{-\ha}_\alpha+\frac12\delta^\ha_\hb
\theta^{+\alpha}_\hc \partial^{-\hc}_\alpha
+\bar\theta^{-\ha\dot\alpha} \bar\partial^{+}_{\hb\dot\alpha}-\frac12\delta^\ha_\hb
\bar\theta^{-\hc\dot\alpha} \bar\partial^{+}_{\hc\dot\alpha}.
\eea

The integral measure in the analytic superspace  has the form
\bea
&&d\zeta^{(-8)}=d^4x_\A D^{-4}\bar{D}^{-4}
,\\
&&D^{-4}=\frac1{24}(D^{-\ha}D^{-\hb})(D^-_{\ha}D^-_{\hb}),\quad\bar{D}^{-4}=\frac1{24}
(\bar{D}^{-a}\bar{D}^{-b})(\bar{D}^-_a\bar{D}^-_b),\nn
&&D^{-4}\Theta^{+4}=1,\quad \bar{D}^{-4}\bar\Theta^{+4}=1.
\eea

\subsection*{B.2~ Superconformal transformations in analytic basis}

The even and odd parameters of the $N=4$ superconformal group  $SU(2,2|4)$ are
\bea
&&b,\quad \omega^\alpha_\beta,\quad \lambda^k_l,\quad
 \epsilon^\alpha_k,\quad \bar\epsilon^{k\dot\alpha},\quad \eta^{k\alpha},\quad
\bar\eta^{\dot\alpha}_k.
\eea
We start from the superconformal transformations of harmonics
\bea
&&\delta_{sc} u^{+a}_l=-\Lambda^{++a}_\hb u^{-\hb}_l,\quad \delta_{sc} \bar{u}^{+k}_\ha
=-(\delta_{sc} u^{+a}_l)^\sim=\Lambda^{++b}_\ha \bar{u}^{-k}_b,\\
&&\delta u^{-\ha}_k=0,\quad \delta \bar{u}^{-k}_\ha=0
\eea
where the composite parameter
\bea
\Lambda^{++a}_\ha=2i\theta^{+\beta}_\ha\bar\theta^{+a\dot\beta}k_{\beta\dot\beta}
+2i\bar{u}^{+k}_\ha u^{+a}_l\lambda^l_k
-2i\bar\theta^{+a\dot\beta}\bar{u}^{+k}_\ha\bar\eta_{k\dot\beta}
+2i\theta^{+\beta}_\ha u_k^{+a}\eta^k_{\beta}=-(\Lambda^{++a}_\ha)^\sim
\eea
satisfies the conditions
\be
D^{++b}_\hb \Lambda^{++a}_\ha=0,\quad D^c_b\Lambda^{++a}_\ha=\delta^a_b\Lambda^{++c}_\ha-\frac12\delta^c_b \Lambda^{++a}_\ha,\quad
\hat{D}^\hc_\hb\Lambda^{++a}_\ha=-\delta^\hc_\ha\Lambda^{++c}_\ha+\frac12\delta^\hc_\hb \Lambda^{++a}_\ha.
\ee
These transformations preserve the basic relations for harmonics \p{A4}-\p{A7}. The
superconformal transformations of the analytic coordinates have the form
\bea
&&\delta_{sc}\,\theta^{+\alpha}_\ha=\frac12b\theta^{+\alpha}_\ha+\omega^\alpha_\beta\theta^{+\beta}_\ha
 +\theta^{+\beta}_\ha k_{\beta\dot\beta}x_A^{\dot\beta\alpha}
-2i\bar{u}^{+l}_\ha u_j^{-\hb}\theta^{+\alpha}_\hb\lambda^j_l\nn
&&+\epsilon^\alpha_k\bar{u}^{+k}_\ha+\bar{u}^{+j}_\ha x_A^{\dot\beta\alpha}\bar\eta_{j\dot\beta}
+2i\theta^{+\beta}_\ha \theta^{+\alpha}_\hb u_j^{-\hb}\eta^j_\beta,\\
&&\delta_{sc}\,\bar\theta^{+a\dot\alpha}=\frac12b\bar\theta^{+a\dot\alpha}+\bar\omega^{\dot\alpha}_{\dot\beta}
\bar\theta^{+a\dot\beta}
 + \bar\theta^{+a\dot\beta}k_{\beta\dot\beta}x_A^{\dot\alpha\beta}
+2i\bar{u}^{-j}_b u_l^{+a}\bar\theta^{+b\dot\beta}\lambda_j^l\nn
&&+\bar\epsilon^{k\dot\alpha}u^{+a}_k+u^{+a}_j x_A^{\dot\alpha\beta}\eta^j_{\beta}
-2i \bar\theta^{+b\dot\alpha}\bar\theta^{+a\dot\beta} \bar{u}_b^{-j}\bar\eta_{j\dot\beta}
=-(\delta_{sc}\,\theta^{+\alpha}_\ha)^\sim,\\
&&\delta_{sc}\,x^{\dot\alpha\alpha}_A=c^{\dot\alpha\alpha}+b x^{\dot\alpha\alpha}_A+\omega^\alpha_\beta x^{\dot\alpha\beta}_A
+\bar\omega^{\dot\alpha}_{\dot\beta} x^{\dot\beta\alpha}_A+x_A^{\dot\alpha\beta}k_{\beta\dot\beta}x_A^{\dot\beta\alpha}
-4\theta^{+\alpha}_\hb\bar\theta^{+b\dot\alpha}\bar{u}^{-j}_bu_k^{-\hb}\lambda^k_j\nn
&&+2i\theta^{+\alpha}_\hb\bar\epsilon^{k\dot\alpha}u_k^{-\hb}
+2i\bar\theta^{+b\dot\alpha}\epsilon^\alpha_k\bar{u}^{-k}_b
+2i\bar\theta^{+b\dot\alpha}\bar{u}^{-j}_b\bar\eta_{j\dot\beta}x^{\dot\beta\alpha}_A
+2i\theta^{+\alpha}_\hb u_j^{-\hb}x^{\dot\alpha\beta}_A\eta^j_\beta.
\eea
Using the complex conjugation of harmonics \p{I1} and the corresponding conjugation of other AB
coordinates
\bea
&&\overline{\theta^{+\alpha}_\ha}=\bar\theta^{-\ha\dot\alpha},\quad \overline{\bar\theta^{+a\dot\alpha}}
=\theta^{-\alpha}_a,\quad \overline{x^m_A}=x^m_A+2i(\theta^-\sigma^m\bar\theta^{+a})-2i(\theta^{+}_\ha\sigma^m\bar\theta^{-\ha})
\eea
we can obtain  superconformal transformations of the spinor coordinates with negative charges, for instance,
\bea
&&\delta_{sc}\,\theta^{-\alpha}_a=\overline{\delta_{sc}\,\bar\theta^{+a\dot\alpha}}=\frac12b\theta^{-\alpha}_a+\omega^{\alpha}_{\beta}
\theta^{-\beta}_a
+\theta^{-\beta}_ak_{\beta\dot\beta}[x_A^{\dot\beta\alpha}+2i\theta^{-\alpha}_a\bar\theta^{+a\dot\beta}
-2i\theta^{+\alpha}_\ha\bar\theta^{-\ha\dot\beta}]\nn
&&
-2iu^{+b}_j\bar{u}^{-l}_a \theta^{-\beta}_b\lambda^j_l+\epsilon^{\alpha}_k\bar{u}^{-k}_a+\bar{u}^{-j}_a[ x_A^{\dot\alpha\beta}+2i\theta^{-\beta}_a\bar\theta^{+a\dot\alpha}
-2i\theta^{+\beta}_\ha\bar\theta^{-\ha\dot\alpha}]\bar\eta_{j\dot\beta}\nn
&&
+2i\theta^{-\beta}_a \theta^{-\alpha}_bu^{+b}_j\eta^j_{\beta}.
\eea
We obtain the superconformal transformations of the harmonic derivatives
\bea
&&\delta_{sc}D^{++a}_\ha=\Lambda^{++a}_\hb D^\hb_\ha-\Lambda^{++b}_\ha D^a_b-\frac12\Lambda^{++a}_\ha D^0,\\
&&\delta_{sc}D^{--\hb}_b=-(D^{--\hb}_b\Lambda^{++c}_\hc)D^{--\hc}_c,\quad
\delta_{sc}[D^{++a}_\ha, D^{--\hb}_b]=0.
\eea

\end{document}